\documentclass[aps,eqsecnum,nofootinbib,superscriptaddress,showpacs]{revtex4}
\def\be{\begin{eqnarray}}
\def\ee{\end{eqnarray}}

\def\p{\partial}

\newcommand{\Exp}[1]{\left\langle~#1~\right\rangle}
\usepackage{amsmath,amsfonts,amssymb,bm}
\usepackage{wrapfig}
\usepackage{graphicx}
\usepackage[dvips]{color}
\begin{document}
\title{Thermalization of boosted charged AdS black holes by an ionic Lattice}  
\date{\today} 

\begin{abstract}
We investigate thermalization process of boosted charged AdS black holes in the Einstein-Maxwell system 
in the presence of an ionic lattice formed by spatially varying chemical potential. 
We calculate perturbations of the black holes by the lattice and investigate how the momentum relaxation 
occurs through umklapp scattering. 
In the WKB approximation, both of the momentum relaxation rate and entropy production rate are analytically obtained and the first law of black holes is 
derived in the irreversible process. 
Interestingly, both the analytical and numerical calculations show that the momentum relaxation rate or the entropy production rate 
does not approach zero in the zero temperature limit unless the velocity of the black hole is zero. 
In the dual field theory side, this indicates that persistent current
 does not exist even in the zero temperature limit, implying 
that the ``ionic lattice'' does not behave as a perfect lattice in a strongly coupled dual field theory.     
\end{abstract}

\author{Akihiro Ishibashi}
\email{akihiro@phys.kindai.ac.jp}
\affiliation{Department of Physics, Kinki University, Higashi-Osaka, 
577-8502, Japan}

\author{Kengo Maeda} 
\email{maeda302@sic.shibaura-it.ac.jp}
\affiliation{Faculty of Engineering,
Shibaura Institute of Technology, Saitama, 330-8570, Japan}

\maketitle

\section{Introduction}
The AdS/CFT correspondence~\cite{adscft} is a useful tool to study strongly correlated condensed matter systems using a classical 
theory of general relativity in asymptotically AdS spacetime. 
For example, charged AdS black holes dressed with a complex scalar field have been investigated as a holographic gravity model of 
a superfluid or a superconductor~(See Ref.~\cite{horowitz_lecture} for a nice review of holographic superconductor models). 
Although typical phenomena of superconductors such as energy gap are explained by the gravity model, there was a serious defect in the model; 
there is no mechanism of momentum dissipation, as translational invariance is imposed. 
This implies that even in the non-superconducting state,  current is conserved and the DC-conductivity becomes infinite.   

In condensed matter systems, momentum conservation is not generally
satisfied via umklapp scattering process since translational invariance
is broken by the presence of an ionic lattice or impurity. 
To incorporate the dissipation mechanism in the holographic model, 
charged AdS black holes with no translational symmetry has been constructed 
perturbatively or numerically in the non-superconducting 
state~\cite{MaedaOkamuraKoga, HorowitzSantosTong2012_1, HorowitzSantosTong2012_2}. 
In Refs.~\cite{HorowitzSantosTong2012_1, HorowitzSantosTong2012_2}, conductivity in the normal phase has been 
investigated and it was shown that the Drude peak appears instead of delta function at zero frequency. This indicates that the holographic model 
yields finite DC-conductivity in the normal phase. In the superconducting state, 
charged hairly AdS black hole solutions with an ionic lattice has been constructed and it has been shown that 
delta function appears, indicating the existence of a superfluid component~\cite{HorowitzSantos2013}. 
This agrees with an earlier result derived from a toy model of
holographic superconductors, showing that the delta function is stable
against adding a lattice~\cite{IizukaMaeda}.   
   
The appearance of the superfluid component shown in
Ref.~\cite{HorowitzSantos2013,IizukaMaeda} predicts the existence of
``persistent current''in the holographic superconductor model with no
translational symmetry. In condensed matter systems, it appears as a
result of macroscopic quantum coherence state and the ``persistent current'' 
does not decay~(at least within the age of our universe).   
So, it is quite interesting to construct such a black hole solution in AdS with ``persistent current'' 
to realize a macroscopic quantum coherence state in the classical theory of general relativity 
dual to the superconducting state. Even in the normal~(non-superconducting) state, there are several results indicating that there should be ``persistent current'' in 
the zero temperature limit~\cite{HartnollHofman2012, HorowitzSantosTong2012_1, HorowitzSantosTong2012_2}.  
Ref.~\cite{HartnollHofman2012} analytically showed that momentum relaxation rate is given as a function 
of a positive power of temperature by computing retarded Green's functions at low temperature and 
frequencies for a static black hole in $AdS_2\times {\bf R}^2$~(See also \cite{DonosHartnoll2013} for a different 
system from the present one, in which there is a static black hole solution that corresponds to an insulating phase 
in which DC-conductivity decreases as temperature goes to zero.). 
%
Refs.~\cite{HorowitzSantosTong2012_1, HorowitzSantosTong2012_2} numerically estimated 
the relaxation time from the Drude form and found that it becomes infinite as temperature goes to 
zero in the presence of a non-perturbative lattice. 
These results suggest that if momentum or current is initially given, it does not decay in the 
zero temperature limit even in the presence of an ionic lattice. 
In other words, there should be extremal charged stationary black hole solutions where 
horizon has nonzero velocity with respect to the lattice, corresponding to 
the existence of ``persistent current'' in the dual field theory side.  

{From the perspective of classical general relativity, 
the rigidity theorem~\cite{HawkingEllis,HollandsIshibashiwald2007} 
guarantees the existence of an additional symmetry
for such a stationary black hole: It states that
if a stationary black hole is non-static and the horizon cross-sections
are compact, then there must be a one-parameter cyclic isometry group
which is spacelike near infinity\footnote{Quite recently, a stationary black hole solution with no such a killing orbit was 
numerically found. The horizon, however, is non-compact, evading the rigidity theorem~\cite{FiguerasWiseman}. }. 
Note that in order to prove the theorem, the analyticity
is required. 
In the presence of a lattice, it is clear that such a Killing
symmetry which would correspond to the translational symmetry for 
the planar horizon case does not exist if the analyticity on the AdS
black holes is imposed. 
So, it agrees with the observation in the dual field theory side 
that initial current or momentum decays according to the finite 
DC resistivity~(or finite momentum relaxation rate) in finite temperature case. 
Even in the zero temperature case, a version of the rigidity theorem still
holds~\cite{HollandsIshibashi2009} and no counter-example
has been found until now. This implies that unless analyticity on the horizon
(or other relevant technical conditions) is violated, one would not be able to
construct an extremal black hole with ``persistent current.''}

Although there have been several studies~\cite{HartnollHofman2012, HorowitzSantosTong2012_1, 
HorowitzSantosTong2012_2, HorowitzSantos2013} on DC resistivity 
by perturbing static charged AdS black holes, little is known about the nature of charged AdS black holes with momentum or current. 
Motivated by this, we investigate how initial momentum or current of 
the AdS black holes decays in the presence of an ionic lattice in the normal 
state.  In the absence of a lattice, charged AdS black hole solutions with ``persistent current'' 
can be constructed by boosting planar Reissner-Nordstr\"{o}m AdS black hole solution. 
In this paper, as a first step, we investigate the stability of the black holes 
against a perturbed ionic lattice and how thermalization occurs by losing the momentum. 
In addition, we examine if the extremal black hole with ``persistent current'' exists in the 
zero temperature limit, avoiding the apparent
inconsistency between no translational symmetry due to the lattice
structure and the consequence of a Killing symmetry due to the rigidity
theorem for stationary black holes. 
The perturbation is given by adding a small spatially varying chemical potential, corresponding 
to the spatially varying gauge potential at infinity. 
At first order in the perturbation, there is no dissipation and we can construct charged AdS black hole solutions 
with ``persistent current'' in the presence of the ionic lattice. At second order in the perturbation, however, 
dissipation occurs and the momentum is lost by the lattice. 

We conduct perturbation analysis by exploiting the gauge invariant
formalism developed by Kodama-Ishibashi~\cite{KodamaIshibashi2004}. 
We first numerically calculate the momentum relaxation rate from the two-decoupled scalar mode equation and 
investigate the dependence of the temperature and the velocity of the black hole. 
For a fixed temperature, the momentum relaxation rate is 
proportional to the velocity of the black hole and it does not decay to zero in the zero temperature limit. 
We next obtain the analytical expression of the momentum relaxation rate
and the entropy production rate in the WKB formalism in the large limit
of the wave number of the lattice. We also show that the physical process version of the first law is satisfied in the irreversible process and 
that thermalization always occurs by losing the momentum, being independent
of the temperature. 

The organization of the paper is as follows. We first derive two master equations of the scalar mode 
perturbations and mention to the boundary conditions in Sec.~II. In Sec.~III, we numerically 
calculate the momentum relaxation rate for various temperature and wave
numbers. In Sec.~IV, we obtain an analytical expression of the momentum
relaxation rate in the WKB formalism. We also calculate the entropy
production rate under the WKB approximation and derive the 
physical process version of the first law of the black hole 
in the irreversible process in Sec.~V. 
Sec.~VI. is devoted to conclusion and discussions. 
\section{Perturbations of boosted Reissner-Nortstr\"{o}m AdS black holes}
In this section, we consider perturbations of boosted Reissner-Nortstr\"{o}m AdS black holes by adding an ionic 
lattice on the boundary theory. As realized in Ref.~\cite{MaedaOkamuraKoga}, it can be constructed by considering 
a spatially varying chemical potential on the boundary theory. According to the AdS/CFT correspondence~\cite{adscft}, 
the chemical potential $\mu$ is given by the boundary value of the
temporal component of the gauge potential, $A_t$. 
So, this corresponds to the scalar-type perturbations in the classification of gauge-invariant 
perturbations~\cite{KodamaIshibashi2004}.     

In the following subsection A, we give the perturbed equations and derive two master variables, 
following~\cite{KodamaIshibashi2004}. In the subsection B, the boundary conditions for the perturbations are summarized.      

\subsection{perturbed equations}
We begin with the following 
four-dimensional Einstein-Maxwell system with action 
\begin{align}
\label{action}
S=\int d^4x\sqrt{-g}\left(R+\frac{6}{L^2}-\frac{1}{2}F^2\right),  
\end{align}
where $L$ is the AdS radius. 
In the absence of the lattice, we are interested, as our background black hole, 
the following boosted planar Reissner-Nortstr\"{o}m AdS black hole solution given by 
\begin{align}
\label{sol:boosted_BH}
& ds^2=-\left(f\cosh^2\beta-\frac{r^2}{L^2}\sinh^2\beta\right)d{\hat{t}_\ast}^2+\frac{dr^2}{f}
+2\left(fL-\frac{r^2}{L}\right)\sinh\beta\cosh\beta d\hat{t}_\ast d\hat{x}_\ast \nonumber \\
&\quad +(r^2\cosh^2\beta-fL^2\sinh^2\beta){d\hat{x}_\ast}^2+r^2d\hat{y}^2, 
\qquad f(r)=\frac{r^2}{L^2}-\frac{2M}{r}+\frac{Q^2}{r^2},   
\end{align}
where $\beta$ is a boost parameter. The null geodesic generator $l^\mu$~\footnote{The freedom for multiple scaling is fixed in another coordinate system adopted in Sec.~III} on the horizon and the 
surface gravity $\kappa_*$ associated with $l^\mu$ are  
\begin{align}
\label{generator_temperature}
l=L\p_{\hat{t}_*}+\tanh\beta\, \p_{\hat{x}_*}, \qquad 
\kappa_*=\frac{L}{2\cosh\beta}\,f_{,r}(r_+)=\frac{L}{\cosh\beta}\left(\frac{3r_+}{2L^2}-\frac{Q^2}{2r_+^3}\right), 
\end{align}
where $r_+$ is the radius of the horizon satisfying $f(r_+)=0$. 

We are primarily interested in considering static perturbations of the following type,  
\begin{align}
\label{static_lattice_perturbation}
\delta A_t(r,\hat{x}_*)=\epsilon a_t(r) e^{-i\hat{k}_*\hat{x}_*}, 
\end{align}
on the boosted black hole background
(\ref{sol:boosted_BH}), where here and hereafter $\epsilon$ denotes
an infinitesimally small parameter and $\hat{k}_*$ is
a wave number of the lattice. However, we do not attempt to
calculate such static perturbations directly on the above boosted black
hole background, as they are complicated on the frame $(\hat{t}_*,\,\hat{x}_*)$.
Instead, we introduce another inertial frame, $(\hat{t},\,\hat{x})$
by the boost \begin{align}
\label{boost_relation}
{\hat{t}/L \choose \hat{x}}=
\left(\begin{array}{rr}
\cosh\beta & -\sinh\beta \\
-\sinh\beta & \cosh\beta \\
\end{array} \right)
{\hat{t}_\ast/L \choose \hat{x}_\ast},  
             \end{align}
and consider relevant perturbations in this new frame. 
By doing so, we can exploit the established pertubation
formulas for static black holes~\cite{KodamaIshibashi2004},
as in this new frame $(\hat{t},\,\hat{x})$, our background
black hole becomes static~\footnote{Strictly speaking, the frame $(\hat{t},\,\hat{x})$ 
is not inertial frame because $\beta$ is not constant during the momentum relaxation process. 
However, the process is almost adiabatic as we consider perturbation and hence we treat it as almost constant.}. 
Hereafter, we shall call the former frame
$(\hat{t}_\ast,\,\hat{x}_\ast)$ and the latter frame
$(\hat{t},\,\hat{x})$ the {\it static lattice frame} and the {\it static black hole frame}, respectively.  

The static perturbation~(\ref{static_lattice_perturbation}) corresponds to the time-dependent perturbation with $e^{i(\hat{k}\hat{x}-\hat{\omega} \hat{t})}$ in the 
the static black hole frame where $\hat{k}$ and $\hat{\omega}$ are given by 
\begin{align}
\label{boost_wave_number}
\hat{\omega}=\frac{\hat{k}_*}{L}\sinh\beta, \qquad \hat{k}=-\hat{k}_*\cosh\beta. 
\end{align}
In the black hole static frame, it is convenient to use $(t,\,u,\,x,\,y)$ coordinate system and the corresponding frequency $\omega$, wave number $k$ defined by 
\begin{align}
u=\frac{r_+}{r}, \quad t=\frac{r_+}{L^2}\hat{t}, \quad x=\frac{r_+}{L}\hat{x}, \quad y=\frac{r_+}{L}\hat{y},   
\quad k=\frac{L}{r_+}\hat{k}, \quad \omega=\frac{L^2}{r_+}\hat{\omega}.    
\end{align}
Since the perturbation corresponds to the scalar-type perturbation, by
choosing the gauge as $f_a~(a=t,\,r)=H_T=0$ in Ref.~\cite{KodamaIshibashi2004},
the perturbed metric in the black hole static frame is given by 
\begin{align}
\label{pert_metric}
& ds^2=\frac{L^2}{u^2}\Biggl[-g(u)\{1+\epsilon(X(u)+2H_L(u))e^{-i\omega t+ikx}\}dt^2
+\frac{1-\epsilon(X(u)+2H_L(u))e^{-i\omega t+ikx}}{g(u)}du^2 \nonumber \\
& \,\,\,-\frac{2\epsilon L^2u^2}{r_+^2g(u)}Z(u)e^{-i\omega t+ikx}dtdu
+(1+2\epsilon H_L(u))e^{-i\omega t+ikx}(dx^2+dy^2)\Biggr], \nonumber \\
& g(u)=(1-u)(1-\xi u)(1+(1+\xi)u+(1+\xi+\xi^2)u^2), 
\end{align}
where the black hole horizon is located at $u=1$ and $\xi$ is a non-extremal parameter in the range of  
$0\le \xi\le 1$~($\xi=0$ and $\xi=1$ correspond to the AdS black hole without charge and the extremal black 
hole, respectively.).
The mass and charge density are the functions of $r_+$ and $\xi$,  
\begin{align}
\label{BC_mass_charge}
& M:=\frac{r_+^3}{2L^2}(1+\xi)(1+\xi^2), \nonumber \\
& Q:=\frac{r_+^2}{L}\sqrt{\xi(1+\xi+\xi^2)}   
\end{align}
and the only non-zero component of the background Maxwell field 
 ${\cal F}_{\mu\nu}$ 
is 
\begin{align}
\label{BC_flux}
 {\cal F}_{ut}=\frac{\sqrt{2}L^2}{r_+^2}Q. 
\end{align}

The scalar-type metric perturbation variables, $H_L,\;X, \; Z$,
are gauge invariant in our present gauge choice (see (5.7a), (5.27) of
\cite{KodamaIshibashi2004}), while the scalar-type perturbation of
the Maxwell field, $\delta F_{\mu\nu}$ is given, 
in terms of a function $A(u)$ (see (5.17) of~\cite{KodamaIshibashi2004}), by 
\begin{align}
\label{def:gauge_field}
& \delta F_{tu}=\epsilon\left[\frac{\omega^2}{g}A+(g A')'\right]e^{-i\omega t+ikx}, \nonumber \\
& \delta F_{tx}=i\epsilon kg A' e^{-i\omega t+ikx}, \qquad 
\delta F_{ux}=\epsilon \frac{\omega k}{g}A e^{-i\omega t+ikx}, 
\end{align}
where here and hereafter {\it prime} denotes
the derivative with respect to $u$. 

The perturbed Einstein equations are then reduced to the following three coupled differential equations 
\begin{align}
\label{eq_H}
g^2H_L''- \left( \omega^2 - k^2 g \right) H_L 
       + \frac{k^2}{2}gX -\frac{L^2 \omega^2 u}{r_+^2}\frac{Z}{i\omega} 
 = 0, 
\end{align}
\begin{align}
\label{eq_tu}
4ugH_L'-2(2g+ug')H_L-2gX
- \frac{L^2k^2u^3 }{r_+^2}\frac{Z}{i\omega}=0, 
\end{align}
\begin{align}
\label{eq_tx}
Z'+i\frac{\omega r_+^2}{L^2u^2}X-\frac{2\sqrt{2}iQ\omega}{L^2}A=0, 
\end{align}
where the last two equations correspond to the Hamiltonian and the momentum constraint equations, respectively when evolution in the 
$u$ direction is regarded as ``time'' evolution. The Maxwell equation yields the evolution equation for $A$ as 
\begin{align}
\label{eq_gauge}
g^2A''+gg'A'+(\omega^2-k^2g)A-\frac{2\sqrt{2}L^2Q}{r_+^2}gH_L=0. 
\end{align} 
Following Ref.~\cite{KodamaIshibashi2004}, we shall introduce 
a master variable $\Phi$ as  
\begin{align}
\label{def_masterPhi}
\Phi(u)=\frac{4\omega r_+^2 H_L-2iL^2u Z}{\omega r_+ uh(u)}, 
\end{align}
where $h$ is a function of $u$ defined by 
\begin{align}
\label{def_h}
h(u)=\frac{r_+^2}{L^2}\left(k^2 - \frac{g'}{u}\right) . 
\end{align}
Then, using Eqs.~(\ref{eq_H}), (\ref{eq_tu}), (\ref{eq_tx}), and  (\ref{eq_gauge})
we obtain two coupled differential equations for $\Phi$ and $A$, 
\begin{align}
\label{eq_master_Phi}
& g(g\Phi')'+(\omega^2-V_\Phi)\Phi= \nonumber \\
& \frac{4\sqrt{2}Qg}{L^2r_+h^2u}\{2r_+^2(k^2+4\xi(1+\xi+\xi^2)u^2)g+L^2h(k^2u^2+ug'-2g)\}A, 
\nonumber \\
& V_\Phi=\frac{g}{L^2r_+^2u^2h^2}U_\Phi, \nonumber \\
& U_\Phi=-2r_+^4(k^2+\xi(1+\xi+\xi^2)u^2)uh'g
%
+h(k^2+4\xi(1+\xi+\xi^2)u^2)r_+^4k^2u^2 
\nonumber \\
& \qquad +L^2r_+^2uh(ugh''+(ug'-2g)h'),  
\end{align}
\begin{align}
\label{eq_master_A}
& 
 g(gA')'+(\omega^2-V_A)A 
 +\frac{\sqrt{2}Q}{r_+}g 
  \left\{ 
         g\Phi'+ \left(\frac{h'}{h}g-\frac{k^2}{2}u \right) \Phi 
  \right\} =0 , 
\nonumber \\
& V_A=\frac{8Q^2g^2}{r_+^2h}+k^2g. 
\end{align}

It is easy to check that $X$, $H_L$, and $Z$ are derived from the master variables $\Phi$ and $A$ as 
\begin{align}
\label{H_Phi}
%
&
H_L(u)=-\frac{r_+}{2L^2}
       \left[
            g \Phi' + \left(g\frac{h'}{h}-\frac{k^2u}{2} \right)\Phi 
       \right]
+\frac{2\sqrt{2}Qg}{L^2h}A,   
\end{align}
\begin{align}
\label{X_Phi}
& X(u)=\frac{r_+}{2L^4h}\{L^2uhg'+4r_+^2g(k^2+4u^2\xi(1+\xi+\xi^2))+2L^2g(uh'-h)\}\Phi'
\nonumber \\
&+\,\,\,\Biggl[\frac{r_+}{2L^2gh^2}\{-2g^2hh'+2ug^2h'^2+h^2(2u\omega^2+g(ug''-g'))\}
\nonumber \\
%
&
\,\,\,+\frac{1}{L^4r_+uh^2}(k^2+4\xi(1+\xi+\xi^2)u^2)
\{2r_+^4ugh'-hk^2r_+^4 u^2\}   \Biggr]\Phi 
\nonumber \\
&\,\,\,+\frac{4\sqrt{2}Qug}{L^2h}A'
\nonumber \\
&+\,\,\,\frac{2\sqrt{2}Q}{L^4h^2r_+^2}
\{L^4u^2h^2+2L^2r_+^2(g-k^2u^2)h-2r_+^2g(2r_+^2(k^2+4\xi(1+\xi+\xi^2)u^2)+L^2uh')  \}A,  
\end{align}
\begin{align}
\label{Z_Phi}
Z(u)=\frac{i\omega r_+^3 g}{L^4u}\Biggl[ \Phi'
-\frac{1}{2}\left(\frac{g'}{g}-\frac{2h'}{h}\right)\Phi-\frac{4\sqrt{2}Q}{r_+h}A
      \Biggr].  
\end{align}

Introducing two master variables $\Phi_\pm$ as 
\begin{align}
\label{master_Phi_pm}
& \Phi_\pm=a_\pm(u)\Phi+b_\pm A, \nonumber \\
& a_+(u)=\frac{r_+^2 k^2Q}{2L^2}+\frac{3(1+k^2\delta)MQ}{r_+}u, \nonumber \\
& a_-(u)=6(1+k^2\delta)M-\frac{4Q^2}{r_+}u, \nonumber \\
& b_+=\frac{6}{\sqrt{2}}(1+k^2\delta)M, \nonumber \\
& b_-=-\frac{8Q}{\sqrt{2}}, \nonumber \\
& \delta=\frac{-1+\sqrt{1+\frac{16k^2\xi(1+\xi+\xi^2)}{9(1+\xi)^2(1+\xi^2)^2}}}{2k^2}, 
\end{align}
two coupled Eqs.~(\ref{eq_master_Phi}) and (\ref{eq_master_A}) are reduced to 
the two decoupled equations,  
\begin{align}
\label{eq_master_pm}
& g(g\Phi_\pm')'+(\omega^2-V_\pm)\Phi_\pm=0, \nonumber \\
& V_\pm=\frac{g}{L^2r_+^2b_\pm uh^2}U_\pm, \nonumber \\
& U_\pm=4\sqrt{2}Qg\{(2r_+^3a_\pm (k^2+4u^2\xi(1+\xi+\xi^2))+L^2h(\sqrt{2}Qub_\pm-2r_+a_\pm))\}
\nonumber \\
&\,\,\,+L^2uhr_+(k^2b_\pm hr_++4\sqrt{2}Qa_\pm(k^2u+g')). 
\end{align}
As mentioned in \cite{KodamaIshibashi2004}, $\Phi_-$ and $\Phi_+$ correspond to the gravitational 
and electromagnetic modes, respectively. 
For later convenience, we express $\Phi$ and $A$ by the master variables $\Phi_\pm$ as 
\begin{align}
\label{Phi_A_Phi_pm}
\Phi=\frac{b_-\Phi_+-b_+\Phi_-}{b_-a_+-b_+a_-}, \qquad 
A=\frac{a_+\Phi_--a_+\Phi_+}{b_-a_+-b_+a_-}. 
\end{align}
\subsection{Boundary conditions}
The solutions of the two decoupled second order different equations~(\ref{eq_master_pm}) are 
completely determined by imposing the following four boundary conditions, i.~e.~, 
asymptotic boundary conditions and ingoing boundary conditions on the horizon. 
As the asymptotic boundary conditions, we require 
that the metric asymptotically approaches AdS and the amplitude of the 
electric field along $x_*$-direction, $E_{x_*}$ is independent of $k_*$. 
   
By Eqs.~(\ref{eq_master_pm}), we obtain the asymptotic behaviors of $\Phi$ and $A$ as
\begin{align}
& A(u)\simeq a+O(u), \nonumber \\
& \Phi(u)\simeq \phi_0+\phi_1 u. 
\end{align}
Substituting this into Eqs.~(\ref{H_Phi}), (\ref{X_Phi}), and (\ref{Z_Phi}), 
$H_L$, $X$, and $Z$ are asymptotically expanded as a series in $u$ as 
\begin{align}
\label{asm_H}
H_L(u)\simeq \frac{4a L\sqrt{2\xi(1+\xi+\xi^2)}-(\phi_1k^2+3\phi_0(1+\xi)(1+\xi^2))r_+}
{2k^2L^2}+O(u), 
\end{align}
\begin{align}
\label{asm_X}
X(u)+2H_L(u)\simeq O(u), 
\end{align}
\begin{align}
\label{asm_Z}
Z(u)\simeq -\frac{i\omega r_+^2}{k^2L^4u}\, 
\{4a L\sqrt{2\xi(1+\xi+\xi^2)}-(\phi_1k^2+3\phi_0(1+\xi)(1+\xi^2))r_+\}+O(1). 
\end{align}
Thus, we shall impose
\begin{align}
\label{bc_infinity}
\phi_1=\frac{4La\sqrt{2\xi(1+\xi+\xi^2)}-3\phi_0(1+\xi+\xi^2+\xi^3)r_+}{k^2r_+} 
\end{align}
so that $O(H_L)=O(X)=O(uZ)=u$,
as an asymptotic boundary condition at infinity~($u=0$). 

Another asymptotic boundary condition is concerned with the electric field along $x_*$-direction. 
By requiring that the amplitude of the electric field $E_{x_*}(u=0)=\delta F_{t_*x_*}(u=0)$ be 
independent of $k_*$ and using the relation~(\ref{boost_wave_number}),
we obtain the second asymptotic boundary condition as  
\begin{align}
\label{bc_chemical}
|\delta F_{tx}(u=0)|=|\delta F_{t_*x_*}(u=0)|\sim \epsilon k_*\cosh \beta A'(0)=\epsilon\, C, 
\end{align}
where $C$ is a constant. 

With respect to the boundary condition on the horizon, we impose the
ingoing boundary condition since we require causal propargation on
the perturbation. 
In terms of the tortoise coordinate, $u_\ast$, the boundary condition is
represented by 
\begin{align}
\label{ingoing_bc}
\Phi_\pm\sim e^{i(kx-\omega (t- u_\ast))}\sim e^{ik(x+v(t- u_\ast))}, 
\qquad u_\ast=\int^u_0 \frac{du}{g}, 
\end{align}
where $u_\ast$ is in the range of $0\le u_\ast\le \infty$ and
$v:=\tanh\beta$ is the velocity of the black hole.
 
\section{Numerical calculation of momentum relaxation}
In this section, we numerically calculate the rate of momentum relaxation ${\cal G}$ 
by the ionic lattice and investigate how ${\cal G}$ depends on the 
temperature and the velocity $v$ of the boosted black hole. 
To define the expectation value of the energy-momentum tensor $\Exp{T^{ab}}$ in the dual field theory, 
it is convenient to use the following coordinate system 
\begin{align}
\label{3+1_decomposition}  
& ds^2=N^2 dz^2+\gamma_{ab}(d\tilde{x}^a+N^adz)(d\tilde{x}^b+N^bdz)
 \quad ~(a,\,b=\tilde{t}_\ast,\,\tilde{x}_\ast,\,\tilde{y}),  \nonumber
 \\ 
&\,\,\to \frac{L^2}{z^2}(dz^2+\eta_{ab}d\tilde{x}^ad\tilde{x}^b+O(z^3)), \quad z\to 0, 
\end{align}
where the spacetime is foliated by $z=\mbox{const}$. timelike hypersurfaces homeomorphic to 
the AdS boundary~($z=0$) and $\eta_{ab}={\rm diag}(-1,1,1)$.   
Then, $\Exp{T^{ab}}$ in the dual field theory is defind by~(See Appendix A in Ref.~\cite{MaedaOkamuraKoga})  
\begin{align}
\label{EM_ex_value}
\Exp{T^{ab}}=\lim_{z\to 0}2\left(\frac{L}{z} \right)^5\left(\gamma^{ab}K-K^{ab}
-\frac{2}{L}\gamma^{ab} \right),\end{align}   
where $\gamma_{ab}$ and $K_{ab}$ are the induced metric and the extrinsic curvature 
on each timelike hypersurface, respectively.  
Note that the last term on the r.~h.~s. of Eq.~(\ref{EM_ex_value}) is the holographic counter-term. 
Since the metric~(\ref{3+1_decomposition}) is obtained from Eq.~(\ref{sol:boosted_BH}) by coordinate transformation 
$z=L/r$,\,$\tilde{t}_*=\hat{t}_*/L$,\,$\tilde{x}_*=\hat{x}_*$,\,$\tilde{y}=\hat{y}$,  
$\Exp{T^{ab}}$ of the background spacetime (\ref{sol:boosted_BH}) 
becomes  
\begin{align}
\label{def:energy-momentum}
& \Exp{T^{\tilde{t}_*\tilde{t}_*}}=LM(1+3\cosh2\beta), \qquad \Exp{T^{\tilde{x}_*\tilde{x}_*}}=LM(3\cosh2\beta-1), \nonumber \\
& \Exp{T^{\tilde{t}_*\tilde{x}_*}}=6LM\cosh\beta \sinh\beta, \qquad
  \Exp{T^{\tilde{y}\tilde{y}}}=2LM . 
\end{align}
The electric current $\Exp{J^a}$ is also defined by 
\begin{align}
\label{def:R-current}
\Exp{J^a}=\lim_{z\to 0}\left(\frac{L}{z} \right)^3\sqrt{2}F^{a\mu}n_\mu, 
\end{align}
where $n^\mu$ is a unit normal outward-pointing vector orthogonal to each hypersurface.  
By using $\lim_{z\to 0}n^z=-z/L$ and $F_{z\tilde{t}_*}=\sqrt{2}Q\cosh\beta$, 
we obtain the background value of $\Exp{J^{\tilde{t}_*}}$ as 
\begin{align}
\Exp{J^{\tilde{t}_*}}=-2Q\cosh\beta. 
\end{align}

We find the first law of the thermodynamics for the boosted black holes  
\begin{align}
\label{static_first_law} 
\delta \Exp{T^{\tilde{t}_*\tilde{t}_*}}=T \delta s
 +v\delta \Exp{T^{\tilde{t}_*\tilde{x}_*}}, 
\end{align}
where $T=\kappa_*/2\pi$ is the temperature and $s$ is the entropy
density defined by the horizon area $\Sigma=r_+^2\cosh\beta$ per unit length 
$\Delta \tilde{x}_*=\Delta \tilde{y}_*=1$ as 
\begin{align}
\label{def:entropy_density}
s=4\pi \Sigma=4\pi r_+^2\cosh \beta. 
\end{align}
To derive Eq.~(\ref{static_first_law}), we have used that 
$\delta \Exp{J^{\tilde{t}_*}}=0$ because in our model, the total charge does not change during the evolution. 
Eq.~(\ref{static_first_law}) is simply derived from the series of the background stationary boosted black hole solutions. 
In Sec.~V, we derive the first law in the irreversible process where momentum relaxation occurs by an ionic lattice in the WKB 
approximation. 

The conservation law of  $\Exp{T^{ab}}$ is derived from the constraint equation 
in the bulk,  
\begin{align}
\label{Eq:constraint}
0=-D_b(\gamma^{ab}K-K^{ab})+F^{b\mu}n_\mu {F^a}_b, 
\end{align}
where $D_a$ is the covariant derivative with respect to the induced metric $\gamma_{ab}$. 
Substituting Eqs.~(\ref{EM_ex_value}) and (\ref{def:R-current}) into Eq.~(\ref{Eq:constraint}), we 
obtain 
\begin{align}
\label{conservation_law}
\p_b\Exp{T^{ab}}=\sqrt{2}\Exp{J^b}\lim_{z\to 0}(\eta^{ac}F_{cb}).
\end{align}
Since we are interested in the rate of spatially averaged momentum loss, ${\cal G}$, we shall 
define a spatial average $\overline{A}$ of any quantity $A(x)$ satisfying a periodic boundary 
condition $A(x+L)=A(x)$ as 
\begin{align}
\label{average}
\overline{A}=\frac{\int^{x+L}_x A(x) dx}{L}. 
\end{align}
Then, ${\cal G}$ is derived from Eq.~(\ref{conservation_law}) by substituting $a=\tilde{x}_*$ 
and taking the spatial average:   
\begin{align}
{\cal G}:=\overline{\p_{\tilde{t}_*}\Exp{T^{\tilde{x}_*\tilde{t}_*}}} 
=\overline{\p_b\Exp{T^{\tilde{x}_*b}}} . 
\end{align}
This is in fact a quantity of $O(\epsilon^2)$ because of 
the following reasoning. As $\eta^{\tilde{x}_*\tilde{x}_*}F_{\tilde{x}_*\tilde{t}_*}$ 
is already $O(\epsilon)$, the r.~h.~s. of Eq.~(\ref{conservation_law}) is 
also $O(\epsilon)$ or even higher. Since 
the zeroth order current of $\Exp{J^b}$ is spatially homogeneous and 
$\overline{F_{\tilde{x}_*\tilde{t}_*}}=\overline{\delta F_{\tilde{x}_*\tilde{t}_*}}\sim \epsilon\overline{e^{-ik_*x_*}}=0$, 
${\cal G}$ at $O(\epsilon)$ must be zero, and hence ${\cal G}$ should be  $O(\epsilon^2)$. 
In addition to the boundary condition~(\ref{bc_chemical}), 
we also impose the homogeneous electric field to be zero 
at all orders, i.~e.~, $\overline{F_{\tilde{x}_*\tilde{t}_*}}|_{z=0}=0$, as we are not interested in the 
case where the electric current increases by the homogeneous electric field.  
This implies that ${\cal G}$ at leading order includes only the first order of  $\Exp{J^b}$ and it does not include 
any second order perturbations of the gauge field and the metric, $F^{(2)}$, 
$h^{(2)}$, defined by 
\begin{align}
& F_{\mu\nu}={\cal F}_{\mu\nu}+\epsilon F^{(1)}_{\mu\nu}+\epsilon^2  F^{(2)}_{\mu\nu}+\cdots, \nonumber \\
& g_{\mu\nu}=g^{(0)}_{\mu\nu}+\epsilon h^{(1)}_{\mu\nu}+\epsilon^2 h^{(2)}_{\mu\nu}+\cdots. 
\end{align}
It is also noteworthy that ${\cal G}$ does not include any first order perturbations of the metric because 
the metric is asymptotically AdS, as described 
in Eq.~(\ref{3+1_decomposition}).   
%
So, one obtains 
\begin{align}
\label{momentum_loss_rate}
& {\cal G} 
=2\,\overline{F_{\tilde{t}_*z}F_{\tilde{x}_*\tilde{t}_*}} 
=2\,\overline{({\cal F}_{\tilde{t}_*z}+\delta F_{\tilde{t}_*z})\delta F_{\tilde{x}_*\tilde{t}_*}}
\nonumber \\
&=2\,\overline{\delta F_{\tilde{t}_*z}\delta F_{\tilde{x}_*\tilde{t}_*}} \nonumber \\
&=-\frac{2r_+^4}{L^4}\,\overline{\delta F_{t_*u}\delta F_{t_*x_*}}. 
\end{align}
%
This indicates that momentum relaxation does not occur 
at linear order but, it does at the second order, $O(\epsilon^2)$. To derive 
the last equation in Eq.~(\ref{momentum_loss_rate}), 
we have used coordinate transformation $z=uL/r_+$, $\tilde{t}_*=Lt_*/r_+$, 
$\tilde{x}_*=Lx/r_+$, and $\tilde{y}=Ly/r_+$.  
\begin{figure}
 \begin{center}
  \includegraphics[width=7truecm,clip]{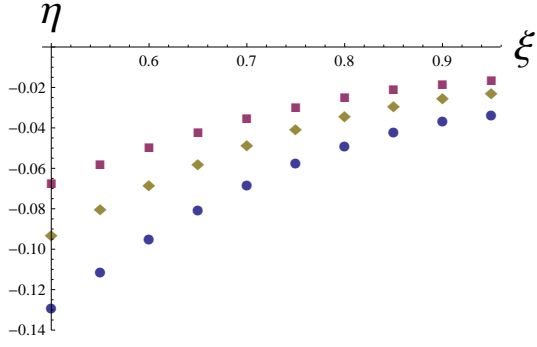}
  \caption{(color online) $\eta=L^4{\cal G}/(\epsilon^2 r_+^4)$ 
is plotted for $k=1/2$ for various $\beta$. 
$\beta=0.1$, $\beta=0.07$, and $\beta=0.05$ correspond to a circle, a rhombus, and 
a square, respectively.} 
 \end{center}
\end{figure}
\begin{figure}
 \begin{center}
  \includegraphics[width=7truecm,clip]{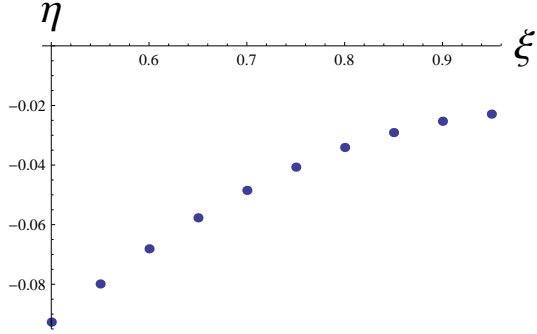}
  \caption{(color online) $\eta:=L^4{\cal G}/(\epsilon^2 r_+^4)$ 
is plotted for $k=1$  and $\beta=0.1$. } 
 \end{center}
\end{figure}
In the static lattice coordinate, $\delta F_{t_*u}$ and $\delta F_{t_*x_*}$ are 
derived from Eq.~(\ref{def:gauge_field}) by the boost Eq.~(\ref{boost_relation}) as 
\begin{align}
\label{boosted_gauge}
& \delta F_{ut_*}=-\epsilon \left[\cosh\beta \left(\frac{\omega^2 A}{g}+(gA')' \right)+\sinh\beta \frac{\omega k}{g}A\right]e^{-ik_*x_*}
=\epsilon (c_R+ic_I)e^{-ik_*x_*}, 
\nonumber \\
 & \delta F_{t_*x_*}=\delta F_{tx}=i\epsilon k g A'  e^{-ik_*x_*}=\epsilon (d_R+id_I)e^{-ik_*x_*},  
\end{align}
where $c_R$, $c_I$, $d_R$, and $d_I$ are some real functions of $u$. 
Then, taking the real value of $\delta F_{t_*u}$ and that of
$\delta F_{t_*x_*}$ in calculating the spatial average of Eq.~(\ref{momentum_loss_rate}), we obtain 
\begin{align}
\label{momentum_loss_rate_N}
{\cal G}=-\frac{2r_+^4}{L^4}\,\overline{\mbox{Re}[\delta F_{t_*u}]\mbox{Re}[\delta F_{t_*x_*}]}=
\frac{\epsilon^2 r_+^4}{L^4}\lim_{u\to 0}(c_Rd_R+c_Id_I). 
\end{align}

We numerically solve the two decoupled Eqs.~(\ref{eq_master_pm}) under the boundary conditions,
(\ref{bc_infinity}), (\ref{bc_chemical}), and (\ref{ingoing_bc}) from
$u=1$ to $u=0$ and evaluate Eq.~(\ref{momentum_loss_rate_N}). To
minimize numerical errors, we replace the second derivative $A''$ in
Eq.~(\ref{boosted_gauge}) by the first ones by Eq.~(\ref{eq_master_A}). 

In Fig.~1 we show the velocity dependence $v$~($=\tanh\beta$) of ${L^4\cal G}/r_+^4 \epsilon^2$ 
normalized by $C=1$ in Eq.~(\ref{bc_chemical}) for $k=1/2$. 
As expected in condensed matter systems, ${\cal G}$ is proportional to $\beta\simeq v$ 
for any $\xi$ when $\beta$ is small. This indicates that in the dual field theory, the equation of motion for 
momentum dissipation is given by 
\begin{align}
\label{effective_dissipation_law}
\frac{d\overline{\Exp{T^{\tilde{x}_*\tilde{t}_*}}}}{dt_*}\simeq-\gamma v
\end{align}
in the presence of an ionic lattice, where $\gamma$ is a function of temperature $T$ and the horizon radius $r_+$. When $v$ is small enough compared with the velocity of light, 
$v\ll 1$, the expectation value of the background spacetime becomes 
$\Exp{T^{\tilde{x}_*\tilde{t}_*}}\sim v$ by Eq.~(\ref{def:energy-momentum}). 
Since we consider perturbations around the expectation value, we obtain 
$\overline{\Exp{T^{\tilde{x}_*\tilde{t}_*}}}\simeq \Exp{T^{\tilde{x}_*\tilde{t}_*}}\sim v$. 
Substituting this into Eq.~(\ref{effective_dissipation_law}), 
we find conventional behavior of dissipation observed in condensed matter systems, 
\begin{align}
\label{time_evolution}
v\sim v_0\,e^{-\frac{t}{\tau}}, 
\end{align} 
where $v_0$ is the initial velocity of the black hole and $\tau$ is the relaxation time determined 
by the amplitude of the lattice $\epsilon$, wave number $k_*$, and temperature and so on.  

In Fig.~2, we show the normalized $\eta=L^4{\cal G}/r_+^4 \epsilon^2$ at smaller wave length, $k=1$ for $\beta=0.1$, 
where the $\xi$ dependence of $\eta$ is the same as the one in Fig.~1. In either case, 
we find that the rate of momentum loss does not approach zero in the extremal limit, $\xi\to 1$, 
independent of the parameters, $\beta$ and the wave number $k$. This implies that there is 
no stationary charged AdS black holes with ``persistent current'' even in the zero temperature 
limit in the presence of ionic lattice. In other words, the ionic lattice 
cannot behave as a perfect lattice with no dissipation in the zero temperature limit.   
At first glance, this might appear to be a contradiction to 
the results~\cite{HorowitzSantosTong2012_1, HorowitzSantosTong2012_2, HartnollHofman2012}, which state that the dissipation 
disappears in the zero temperature limit. 
Actually, there is no discrepancy between the present analysis and the previous result, since in terms of perturbation, 
the order of effects considered are essentially different between the two:
%
Although the present analysis includes the lattice effect 
as a perturbation, we take into account the non-zero current 
already at zeroth order as an initial state and calculate 
the momentum relaxation rate at non-linear order. 
In contrast, in Refs.~\cite{HorowitzSantosTong2012_1, HorowitzSantosTong2012_2}, 
the lattice effect is included non-linearly, but the current is not considered 
at the background level and the results concern, in essence, linear response induced by a small electric 
field~\footnote{We thank S.~A.~Hartnoll for discussion.}. 
  
Let us consider the time evolution of the charged black hole with initial momentum in the zero temperature limit. Due 
to the non-zero momentum relaxation rate at the initial state, ${\cal G}\neq 0$ at $\xi=1$, the total momentum would 
be lost gradually and the entropy of the black hole should increase because it is an irreversible process. In other 
words, the black hole necessarily heats up even though initially we start from $\xi=1$.  
One might wonder if the perturbed solutions with $\xi\to 1$ is indeed zero temperature solution because it is already perturbed 
by an ionic lattice. By the perturbation, the temperature would slightly change to $O(\epsilon)$, as 
the amplitude of the perturbation is $O(\epsilon)$. However, the effect only appears at higher order corrections 
for ${\cal G}$, i.~e.~, $O(\epsilon^3)$ or even higher because ${\cal G}$ is already $O(\epsilon^2)$ for any temperature. 
So, the fact that ${\cal G}\neq 0$ at $\xi=1$ is a little bit surprising because the thermal fluctuations go to 
zero and the umklapp scattering must disappear in the extremal limit, $\xi\to 1$ unless a residual resistance remains. 
We will discuss the irreversible process in Sec.~V. 
      
\section{WKB analysis of momentum relaxation}
In this section, we support the numerical results in the previous section by 
deriving the momentum relaxation rate ${\cal G}$ analytically in the large limit of 
wave number $k$ by using WKB approximation. The analytic expression of ${\cal G}$ 
will explain the reason why it does not approach zero in the zero 
temperature limit. 
 

Rewriting the master variables $\Phi_\pm$ in Eq.~(\ref{master_Phi_pm}) as $\Phi_\pm=e^{kS_\pm}$, we can expand 
$S_\pm$ and the potential $V_\pm$ in Eq.~(\ref{eq_master_pm}) as a series in $1/k$ as 
\begin{align}
\label{series_k}
& S_\pm=S_{0\pm}+\frac{S_{1\pm}}{k}+\frac{S_{2\pm}}{k^2}+\cdots, \nonumber \\
& V_\pm=k^2\left(V_{0\pm}+\frac{V_{1\pm}}{k}+\frac{V_{2\pm}}{k^2}+\cdots\right). 
\end{align}
Then, substitution of Eq.~(\ref{series_k}) into Eq.~(\ref{eq_master_pm}) yields  
the following equations
\begin{align}
\label{eq_WKB}
\left(\frac{dS_{0\pm}}{du_*}\right)^2=V_{0\pm}-v^2, \qquad 
\frac{d^2S_{0\pm}}{du_*^2}+2\frac{dS_{0\pm}}{du_*}\frac{dS_{1\pm}}{du_*}-V_{1\pm}=0, \cdots, 
\end{align}
where $V_{i\pm}~(i=0,\,1)$ is given by 
\begin{align}
\label{series_V}
V_{0\pm}=g(u), \qquad V_{1\pm}=\mp 2\sqrt{\xi(1+\xi+\xi^2)}\,ug(u). 
\end{align}
\begin{figure}
 \begin{center}
  \includegraphics[width=7truecm,clip]{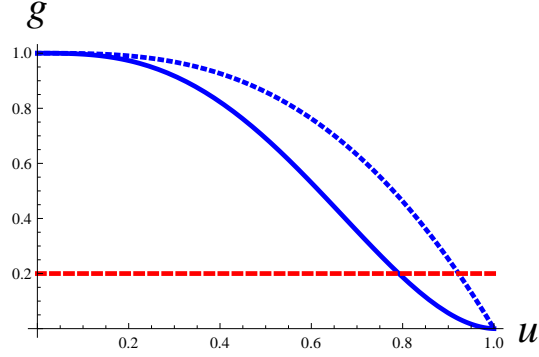}
  \caption{$V_{0\pm}=g$ is shown for $\xi=0.99$~(solid curve) and  $\xi=0.2$~(dotted curve). 
The horizontal line~(dashed line) corresponds to $v^2=1/5$.} 
 \end{center}
\end{figure}

As shown in Fig.~3, there is only one turning point $u_*=u_{*0}$~($v^2-g(u_{*0})=0$) for the potential, 
as $g$ is a monotonously decreasing function of $u$. So, for $u_{*0}<u_*<\infty$, 
the solution of Eq.~(\ref{eq_WKB}) is 
given in terms of ${\cal Q}:=v^2-g$ as 
\begin{align}
\label{sol_WKB>}
& \Phi_-\simeq \frac{D_-}{{\cal Q}^{\frac{1}{4}}}\exp\left(-ik\int^{u_*}_{u_{*0}}\sqrt{{\cal Q}}du_*\right)
\exp\left(i\sqrt{\xi(1+\xi+\xi^2)}\int^{u_*}_{u_{*0}}\frac{ug}{\sqrt{{\cal Q}}}du_*\right) \nonumber \\
&\,\,\,+\frac{D_+}{{\cal Q}^{\frac{1}{4}}}\exp\left(ik\int^{u_*}_{u_{*0}}\sqrt{{\cal Q}}du_*\right)
\exp\left(-i\sqrt{\xi(1+\xi+\xi^2)}\int^{u_*}_{u_{*0}}\frac{ug}{\sqrt{{\cal Q}}}du_*\right), 
\nonumber \\
& \Phi_+\simeq \frac{\tilde{D}_-}{{\cal Q}^{\frac{1}{4}}}\exp\left(-ik\int^{u_*}_{u_{*0}}\sqrt{{\cal Q}}du_*\right)
\exp\left(-i\sqrt{\xi(1+\xi+\xi^2)}\int^{u_*}_{u_{*0}}\frac{ug}{\sqrt{{\cal Q}}}du_*\right) \nonumber \\
&\,\,\,+\frac{\tilde{D}_+}{{\cal Q}^{\frac{1}{4}}}\exp\left(ik\int^{u_*}_{u_{*0}}\sqrt{{\cal Q}}du_*\right)
\exp\left(i\sqrt{\xi(1+\xi+\xi^2)}\int^{u_*}_{u_{*0}}\frac{ug}{\sqrt{{\cal Q}}}du_*\right), 
\end{align}  
and for $0\le u_*\le u_{*0}$ as 
\begin{align}
\label{sol_WKB<}
& \Phi_-\simeq \frac{C_-}{|{\cal Q}|^{\frac{1}{4}}}\exp\left(k\int^{u_{*0}}_{u_{*}}\sqrt{|{\cal Q}|}du_*\right)
\exp\left(\sqrt{\xi(1+\xi+\xi^2)}\int^{u_{*0}}_{u_{*}}\frac{ug}{\sqrt{|{\cal Q}|}}du_*\right) \nonumber \\
&\,\,\,+\frac{C_+}{|{\cal Q}|^{\frac{1}{4}}}\exp\left(-k\int^{u_{*0}}_{u_{*}}\sqrt{|{\cal Q}|}du_*\right)
\exp\left(-\sqrt{\xi(1+\xi+\xi^2)}\int^{u_{*0}}_{u_{*}}\frac{ug}{\sqrt{|{\cal Q}|}}du_*\right), 
\nonumber \\
& \Phi_+\simeq \frac{\tilde{C}_-}{|{\cal Q}|^{\frac{1}{4}}}\exp\left(k\int^{u_{*0}}_{u_{*}}\sqrt{|{\cal Q}|}du_*\right)
\exp\left(-\sqrt{\xi(1+\xi+\xi^2)}\int^{u_{*0}}_{u_{*}}\frac{ug}{\sqrt{|{\cal Q}|}}du_*\right) \nonumber \\
&\,\,\,+\frac{\tilde{C}_+}{|{\cal Q}|^{\frac{1}{4}}}\exp\left(-k\int^{u_{*0}}_{u_{*}}\sqrt{|{\cal Q}|}du_*\right)
\exp\left(\sqrt{\xi(1+\xi+\xi^2)}\int^{u_{*0}}_{u_{*}}\frac{ug}{\sqrt{|{\cal Q}|}}du_*\right).  
\end{align}

Since we require the ingoing boundary condition~(\ref{ingoing_bc}) at the horizon $u_*=\infty$, the coefficients $D_+$ and $\tilde{D}_+$ are set to be zero~(Note that 
$k<0$). Then, the standard connection formulas around the turning point is given by 
\begin{align}
\label{turning_point}
& \tilde{C}_+=-i\tilde{D}_- e^{\frac{\pi i}{4}}, \qquad 
\tilde{C}_-=\frac{1}{2}\tilde{D}_- e^{\frac{\pi i}{4}}, \nonumber \\
& C_+=-iD_- e^{\frac{\pi i}{4}}, \qquad 
C_-=\frac{1}{2}D_- e^{\frac{\pi i}{4}}. 
\end{align}  

By Eqs.~(\ref{Phi_A_Phi_pm}), (\ref{sol_WKB<}), and (\ref{turning_point}), we can expand 
Eq.~(\ref{bc_infinity}) as a series in $1/k$ and $D_-$ can be expressed 
by $\tilde{D}_-$ as, 
\begin{align}
\label{D_tildeD_relation}
D_-=\frac{4L(2\Gamma^2\zeta^2-i)}{r_+ k(2\Gamma^2-i\zeta^2)}\tilde{D}_-+O\left(\frac{1}{k^2} \right),    
\end{align}
where 
\begin{align}
\label{def_Gamma_zeta}
& \Gamma=\exp\left(-k\int^{u_{*0}}_{0}\sqrt{|{\cal Q}|}du_*\right), \nonumber \\
& \zeta=\exp\left(\sqrt{\xi(1+\xi+\xi^2)}\int^{u_{*0}}_{0}\frac{ug}{\sqrt{|{\cal Q}|}}du_*\right). 
\end{align}
$\Phi(0)$ and $A(0)$ are also expressed by $\tilde{D}_-$ as
\begin{align}
\label{def_Phi_A_bc}
& \Phi(0)=-\frac{2e^{\frac{3\pi i}{4}}L^3\Gamma(\zeta^4-1)\tilde{D}_-}
{r_+^4(1-v^2)^{1/4}(2i\Gamma^2\zeta+\zeta^3)\sqrt{\xi(1+\xi+\xi^2)}\,k^2}+O\left(\frac{1}{k^4}\right), 
\nonumber \\
& A(0)=\frac{e^{\frac{3\pi i}{4}}L^2(1+4\Gamma^2)\zeta \tilde{D}_-}
{2\sqrt{2}r_+^3(1-v^2)^{1/4}(2\Gamma^3-i\Gamma\zeta^2)\sqrt{\xi(1+\xi+\xi^2)}k}+O\left(\frac{1}{k^3}\right). 
\end{align}
The unknown coefficient $\tilde{D}_-$ is determined by the boundary condition~(\ref{bc_chemical}) as
\begin{align}
\label{tildeD}
 \tilde{D}_-=\frac{2\sqrt{2}r_+^3e^{-\frac{i\pi}{4}}\Gamma \zeta\, \sqrt{\xi(1+\xi+\xi^2)}\,C}
{L^2k(1-v^2)^{1/4}(1+2i\Gamma^2\zeta^2)}+O\left(\frac{C}{k^2}\right). 
\end{align}
Thus, we obtain 
\begin{align}
\label{Phi_A_k}
\Phi(0)=O(k^{-3}), \qquad A(0)=O(k^{-2}), \qquad \Phi'(0)=O(k^{-4}).  
\end{align}

In the large $k$ limit, ${\cal G}$ is calculated as 
\begin{align}
\label{momentum_loss_rate _WKB}
& {\cal G}=\overline{\p_{\tilde{t}_*}\Exp{T^{\tilde{x}_*\tilde{t}_*}}} \nonumber \\
&=-\frac{2r_+^4}{L^4}\overline{\frac{1}{2}(\delta F_{t_*u}\delta F^*_{t_*x}+\delta F^*_{t_*u}\delta F_{t_*x})}
\nonumber \\
&=-\epsilon^2\frac{r_+^4C^2(\zeta^2+\zeta^{-2})}{L^4\Gamma^2}. 
\end{align}
Here, to obtain the real value of ${\cal G}$, we replaced  $\delta F_{t_*u}\delta F_{t_*x}$ by 
$(\delta F_{t_*u}\delta F^*_{t_*x}+\delta F^*_{t_*u}\delta F_{t_*x})/2$, where $\delta F^*$ is the complex 
conjugate of $\delta F$ and we used the approximation 
\begin{align}
\label{app_A_large_k}
g(gA')'+\omega^2A\Bigl|_{u=0}\simeq V_A(0)A(0)=k^2A(0) 
\end{align}  
in the large $k$ limit. 

As shown in Fig.~3, the potential $V_{0\pm}=g$ at leading order is a monotonously 
decreasing function of $u$ and $g\le 1$. Hence we obtain  
\begin{align}
\ln \Gamma=-k\int^{u_0}_0\frac{\sqrt{|{\cal Q}|}}{g}du<
-k\int^{u_0}_0\frac{\sqrt{|{\cal Q}|}}{v^2}du<-k\int^1_0\frac{du}{v^2}=-\frac{k}{v^2},  
\end{align}
where $u_0$ is the turning point defined by $v^2=g(u_0)$. This indicates that 
$\Gamma$ never diverges as $\xi\to 1$ for any fixed non-zero $v$, and thus 
the rate of momentum loss never goes to zero even in the zero temperature 
limit by Eq.~(\ref{momentum_loss_rate _WKB}). In this sense, the analytical 
result agrees with the numerical results in the previous section.  

It is noteworthy that the energy density does not change during the loss of the 
momentum because we consider static perturbations in the static lattice frame. 
By using the fact 
\begin{align}
\label{deviation_Fux}
& \delta F_{ux_*}(0)=\epsilon \lim_{u\to 0}\left[\sinh\beta \left(\frac{\omega^2 A}{g}+(gA')' \right)
+\cosh\beta \frac{\omega k}{g}A\right]e^{-ik_*x_*}
\nonumber \\
& \simeq \epsilon \lim_{u\to 0}\left[\sinh\beta\, k^2 A+\cosh\beta\, \omega k A\right]e^{-ik_*x_*}=0, 
\end{align}
and substituting $a=t_*$ in Eq.~(\ref{conservation_law}), we can easily 
check 
\begin{align}
\label{Energy_loss_rate}
\overline{\p_{\tilde{t}_*}\Exp{T^{\tilde{t}_*\tilde{t}_*}}}\sim 
\overline{\delta F_{t_*x_*}\delta F^*_{ux_*}+\delta F^*_{t_*x_*}\delta F_{ux_*}}\simeq 0. 
\end{align}  
Here, we used Eq.~(\ref{app_A_large_k}) and $\omega=-k\tanh \beta$ in the second line 
in Eq.~(\ref{deviation_Fux}).
As the total energy is conserved during the relaxation of the momentum, 
the kinetic energy associated with the momentum should be converted into thermal energy via the umklapp scattering 
in the dual field theory. We will address this issue in the next section.

\section{The first law of the black holes in the momentum relaxation process}
In the previous sections, we have seen that momentum relaxation occurs for any temperature 
through the perturbed gauge field $\delta A_\mu$ on the boundary. Since this is an irreversible process, 
entropy should be produced, although the energy density does not change. From the perspective 
of the bulk theory side, both the gravitational and electromagnetic waves are induced from the 
ionic lattice boundary condition~(\ref{bc_chemical}) and fall into the horizon. As the area of the 
black hole increases during the process, the entropy density also does. The process is very similar 
to the Penrose process~(see the text book~\cite{Wald_book}) in the sense that entropy is produced 
in the process that the angular momentum of a rotating black hole is extracted. 
However, one of the key differences from the Penrose process is that energy cannot be extracted from the 
boosted black hole because we consider static perturbations~(in the static lattice frame). Another key difference 
is that the process continues until the momentum of the boosted black hole becomes zero.  

In this section, we calculate the entropy production rate under the WKB approximation, following 
the Hawking-Hartle formula~\cite{Chandra}. We also check that the first law of the black hole is satisfied 
in the irreversible process and that thermalization occurs. 

Let $\rho$ and $\sigma$ respectively be
the expansion and the shear of the null geodesic congruence $l$ along the horizon.
The evolution equations are given by 
\begin{align}
\label{eq_Raychaudhuri}
& l^\mu\p_\mu \rho=\kappa_* \rho-\rho^2-|\sigma|^2-\frac{1}{2}T_{\mu\nu}l^\mu l^\nu, \nonumber \\
& l^\mu\p_\mu \sigma=\kappa_* \sigma+\Psi_0,  
\end{align} 
where $\Psi_0$ is a component of the Weyl tensor in the Newman-Penrose formalism~(see Ref.~\cite{Chandra}) 
and $\kappa_*$ is the surface gravity on the horizon given in Eq.~(\ref{generator_temperature}).  

Since we consider perturbations around the stationary black hole, $\rho$ is very small and hence $\rho$, $\sigma$,  
and $T_{\mu\nu}l^\mu l^\nu$ can be expanded as 
\begin{align}
\label{expansion_rho}
& \rho=\epsilon \rho_{(1)}+\epsilon^2 \rho_{(2)}+\cdots, \nonumber \\
& \sigma=\epsilon \sigma_{(1)}+\cdots, \nonumber \\
& T_{\mu\nu}l^\mu l^\nu=\epsilon {\cal T}_{(1)}+
\epsilon^2{\cal T}_{(2)}+\cdots. 
\end{align} 
Let us define an advanced coordinate $v_*$ as $dv_*=(d\hat{t}-dr/f)\cosh\beta/L$, where it coincides with 
the time coordinate $\tilde{t}_*$ at infinity introduced in Sec.~III.  
Then, $l=\p_{v_*}$ and, as shown in Appendix A, we obtain  
\begin{align}
\label{area_variation}
\frac{d\Sigma}{dv_*}\simeq \frac{2\epsilon^2\Sigma}{\kappa_*}\left(|\sigma_{(1)}|^2
+\frac{1}{2}\overline{{\cal T}_{(2)}} \right),  
\end{align}
where $\Sigma$ is an element of surface of the null congruence on the horizon defined by 
\begin{align}
\label{def_Sigma}
\frac{d\Sigma}{dv_*}=2\rho\Sigma. 
\end{align} 
Since $\Sigma$ is proportional to the entropy density per unit area $\Delta \tilde{x}_*=\Delta \tilde{y}_*=1$, 
Eq.~(\ref{area_variation}) describes the entropy production rate per unit time $\Delta \tilde{t}_*=1$ in the static lattice frame. 

Near the horizon, the component of the Weyl tensor, $\Psi_0$ is
represented by the master variable $\Phi$~(\ref{def_masterPhi}) as 
\begin{align}
\label{Weyl}
& \Psi_0(r_+)=\frac{\epsilon L^2\hat{k}^2 e^{i(kx-\omega t)}(f(2H_L+X)-Z)}{8r_+^2\cosh^2\beta }+O(\epsilon^2) \nonumber \\
& \quad =-\frac{i\epsilon r_+\omega \hat{k}^2(2i\omega-g'(1))}{8L^2\cosh^2\beta}e^{i(kx-\omega t)} \Phi(1)+(\epsilon^2).  
\end{align}
To derive the second equality, we used $g\Phi'\simeq i\omega \Phi$ and the approximation 
\begin{align}
\label{X_Z_horizon}
& X\simeq \frac{r_+}{2L^2}g'\Phi'+\frac{r_+\omega^2}{L^2g}\Phi, \nonumber \\
& Z\simeq \frac{i\omega r_+^3}{L^4}\left(g\Phi'-\frac{1}{2}g'\Phi\right) 
\end{align}
near the horizon. Substituting Eq.~(\ref{Weyl}) into Eq.~(\ref{eq_Raychaudhuri}) and replacing 
$l^\mu\p_\mu\to -iL\hat{\omega}/\cosh\beta$, 
we obtain $\sigma_{(1)}$ as
\begin{align}
\label{sol_shear}
\sigma_{(1)}=\frac{ir_+^2\omega k^2}{4L^3\cosh\beta} e^{i(kx-\omega t)}\Phi.  
\end{align}

From Eqs.~(\ref{master_Phi_pm}) and (\ref{Phi_A_Phi_pm}),  
$\Phi$ reduces to
\begin{align}
\label{Phi_app}
\Phi\simeq-\frac{8Q\Phi_++6k^2\delta M\Phi_-}{\sqrt{2}(b_-a_+-b_+a_-)}
\end{align}
under the WKB approximation. 
Then, at the leading order in $k$, the spatial average of $|\Phi|^2$ becomes 
\begin{align}
\label{Phi_average}
& \overline{|\Phi(1)|^2}=\frac{2L^2C^2}{r_+^2 v \sqrt{1-v^2}\, \Gamma^2 k^6}\left(\frac{1}{\zeta^2}+\zeta^2-2\cos\Theta  \right), \nonumber \\
& \Theta:=2\sqrt{\xi(1+\xi+\xi^2)}\int^\infty_{u_{*0}}\frac{ug}{\sqrt{{\cal Q}}}du_*. 
\end{align} 

In the WKB approximation, the magnitudes of $A$ and $\Phi$ are very small, as $A=O(k^{-2})$, $\Phi=O(k^{-3})$ by Eq.~(\ref{Phi_A_Phi_pm}). 
So, the metric components near the horizon are also small as $X\sim Z\sim H_L=O(k^{-1})$, by Eqs.~(\ref{eq_H}) and (\ref{X_Z_horizon}). 
This implies that the metric fluctuation near the horizon does vanish in the large $k$ limit and the second term in the r.~h.~s. 
of Eq.~(\ref{area_variation}) is simplified to   
\begin{align}
\label{Ricci_average}
& \overline{{\cal T}_{(2)}}=\overline{{T_{\mu\nu}}^{(2)}}l_{(0)}^\mu l_{(0)}^\nu 
\nonumber \\
& =\frac{1}{\epsilon^2}\overline{(\delta F_{tx})(\delta F_{tx})^\ast}\left(\frac{r_+}{L^2}\right)^2\overline{g^{xx}} \nonumber \\
&\quad =\frac{r_+^2}{L^4\cosh^2\beta}\,\overline{|ikg A' e^{-i(\omega t-kx)}|^2}
\nonumber \\
&=\frac{v r_+^2\sqrt{1-v^2}C^2}{4L^4\,\Gamma^2}\left(\zeta^2+\frac{1}{\zeta^2}+2\cos\Theta \right). 
\end{align}
Therefore, by Eqs.~(\ref{area_variation}), (\ref{sol_shear}), (\ref{Phi_average}), and (\ref{Ricci_average}), we finally obtain 
\begin{align}
\label{Area_average}
\frac{d\Sigma}{dv_\ast}
\simeq \frac{\epsilon^2 r_+^2\sqrt{1-v^2}C^2}{2L^4\,\Gamma^2}\left(\zeta^2+\frac{1}{\zeta^2} \right)\frac{v\,\Sigma}{\kappa_*}. 
\end{align}
By Eqs (\ref{generator_temperature}), 
(\ref{def:entropy_density}), (\ref{momentum_loss_rate _WKB}), 
(\ref{Energy_loss_rate}), and (\ref{Area_average}), 
it is easy to check that the first law in the dynamical process is satisfied; 
\begin{align}
\label{first_law}
0=d{\cal E}= Tds+vd{\cal L},  
\end{align}
where ${\cal E}$ and ${\cal L}$ are the energy density $\Exp{T^{\tilde{t}_*\tilde{t}_*}}$ 
and the momentum density $\Exp{T^{\tilde{x}_*\tilde{t}_*}}$ of the background spacetime, respectively.  
This equation means that ${\cal L}$ must decrease to satisfy the second law of the black hole 
thermodynamics, $ds\ge 0$. In other words, the momentum relaxation by the ionic lattice 
is an irreversible process associated with the entropy production.  

The thermalization is guaranteed in the irreversible process as follows. The deviation of the 
temperature $T=\kappa_*/2\pi$ in Eq.~(\ref{generator_temperature}) becomes 
\begin{align}
\label{thermalization}
& dT=-\tanh\beta\,Td\beta+ \frac{L}{2\pi\cosh\beta}\left(\frac{3dr_+}{2L^2}+\frac{3Q^2dr_+}{2r_+^4}
-\frac{QdQ}{r_+^3}  \right) \nonumber \\
&\qquad \ge \frac{L}{2\pi\cosh\beta}\left(\frac{3dr_+}{2L^2}+\frac{3Q^2dr_+}{2r_+^4}
+\frac{Q^2\tanh\beta d\beta}{r_+^3}  \right) \nonumber \\
&\qquad \ge \frac{T}{r_+}dr_+>0. 
\end{align}
Here, we used the facts that $T\ge 0$, $d\beta<0$, and $dQ=-Q\tanh\beta d\beta$, which represents 
$d \Exp{J^{\tilde{t}_*}}=0$, to derive the second inequality. In the third inequality, we used 
$ds=2\pi(4r_+\cosh\beta dr_++2r_+^2\sinh\beta d\beta)\ge 0$.
Eq.~(\ref{thermalization}) means that 
the kinetic energy associated with the initial momentum is converted into thermal energy, and then,  
the temperature increases during the irreversible process.

\section{conclusion and discussions}
We have investigated adiabatic evolution of charged boosted AdS black
holes by perturbation of an ionic lattice. 
At linear order in the perturbation, we constructed charged stationary AdS black hole solutions with 
an ionic lattice. At second order, however, the momentum relaxation occurs by the lattice and the rate 
of momentum loss is proportional to the velocity of the black hole, as shown in Sec.~III. In conventional 
condensed matter systems, the equation of motion for an electron with charge $e$, mass $m$, and velocity $v$ 
is effectively given by 
\begin{align}
m\frac{dv}{dt}=eE-\tilde{\gamma} v, 
\end{align} 
where $E$ is the electric field and $\tilde{\gamma}$ is a positive constant determined by temperature and so on.  
As shown in Eq.~(\ref{effective_dissipation_law}), we have verified that the equation is satisfied in the presence 
of initial velocity when the electric field is zero. The coefficient $\gamma$ in Eq.~(\ref{effective_dissipation_law}) 
corresponding to the coefficient $\tilde{\gamma}$ is a complicated function of temperature, but it never becomes zero 
in the zero temperature limit. This is supported by the analysis of WKB approximation in Sec.~IV. 
This indicates that ``persistent current'' cannot exist
in a condensed matter system that is dual to the present 
charged AdS black hole at zero temperature.  

Even though this result itself is consistent with
the rigidity theorem for stationary black holes, there seems to be an
apparent discrepancy between our result and the previous 
results~\cite{HartnollHofman2012, HorowitzSantosTong2012_1,
HorowitzSantosTong2012_2} which state that DC-conductivity becomes
infinite at zero temperature. One of the reasons is that we calculate non-linear 
perturbations beyond a linear response theory, as mentioned in Sec.~III. Another 
reason is that we take into account the effect of non-zero current at zeroth order in the perturbations. 
In Ref.~\cite{HartnollHofman2012}, the rate of momentum loss can be
obtained
by calculating retarded Green's functions of the perturbed 
gauge potential $\delta A_t\sim e^{-i(\omega t+kx)}$ in the limit of zero frequency, 
$\omega\to 0$ for a static black hole in $AdS_2\times {\bf R}^2$. 
In our setting, the perturbation of the gauge potential $A_t$ is given by 
$\delta A_t\sim e^{-ik_*x_*}$ in the frame where the black hole has
initial momentum or velocity $v$. If we consider the perturbation
in another frame where the velocity of the black hole is zero, 
the form becomes $\delta A_t\sim e^{ik(v t+x)}$~($k<0$), 
implying that the lattice is moving along $x$-direction with constant
velocity $-v$. Thus, in such a static black hole frame,
the perturbations correspond to non-zero frequencies $-kv$,
being different from the perturbation considered
in Ref.~\cite{HartnollHofman2012}.    

In the static black hole frame, energy is always pumped from the boundary into the bulk 
by the moving lattice and then, the 
initially static black hole starts moving until the velocity of the black hole reaches the one 
of the lattice. During the pumping, the energy is always absorbed into the black hole and the 
entropy is produced during the process. This means that thermalization always occurs 
even in the zero temperature limit, as shown in Sec.~V. In the original frame where the black hole has initial momentum 
and velocity, the velocity continues to decrease until it becomes zero, keeping the total energy 
fixed. Since the entropy production is independent of the frame, we have clarified the dissipation 
mechanism of the momentum loss caused by the ``friction'' between the lattice and the velocity 
of the black hole. As a consistency check, we have also derived the first law of black hole in the 
irreversible process.
Although our analysis is limited to the framework of perturbation,
we expect that the same result should also be obtained even
for fully dynamical, non-perturbative case. 
It would be interesting to explore 
non-perturbative dynamics of the present system
by using numerical methods. 
  
There remains several open questions. For example, how can our result be interpreted in 
the dual field theory side? 
Within the framework of our present analysis, the current always decays
even in zero temperature and therefore the lattice cannot be interpreted as a perfect lattice, contrary to the prediction of [4].
One possibility would be that there is a residual resistance in the strongly coupled 
field theory dual to the black hole. The residual resistance
can be caused by impurity or strong interactions between
quasiparticles even in the zero temperature.
It is interesting to explore
our result from the perspective of the dual field theory.
Another open question is whether one can construct black hole solutions
in the presence of ionic lattice dual to a superconducting state
with ``persistent current.''
To reconcile with the symmetry consequence of 
the rigidity theorem, we need to construct a black hole solution with
momentum where the horizon is static with respect to the lattice 
but some condensation of complexed scalar field outside 
the horizon moves along the lattice. 
It would be reported in the near future~\cite{MaedaIshibashiIizuka}.

\section*{Acknowledgments}
We wish to thank Gary T.~Horowitz for valuable discussions. It is also a pleasure to 
acknowledge helpful discussions with S.~A.~Hartnoll. 
K.~M. is supported in part by MEXT/JSPS KAKENHI Grant Number 2374022.
A.I. is supported by the Grant-in-Aid for Scientific
Research Fund of the JSPS (C)No. 22540299 and by the Barcelona Supercomputing Center (BSC) under Grant No. AECT-2012-2-0005.

\section*{Appendix A: The Hawking-Hartle formula}
Integration of the first equation in Eqs.~(\ref{eq_Raychaudhuri}) yields 
\begin{align}
\label{sol1_expansion}
\rho_{(2)}(v_*)=\int^\infty_{v_*} e^{\kappa_*(v_*-v_*')}\left(|\sigma_{(1)}|^2+\frac{1}{2}
{\cal T}_{(2)} \right)dv_*'. 
\end{align}
Substituting Eq.~(\ref{sol1_expansion}) into Eq.~(\ref{def_Sigma}) 
and integrating it by once, we obtain 
\begin{align}
& \ln\left[\frac{\Sigma(v_{*1})}{\Sigma(0)}\right]=
\frac{2\epsilon^2}{\kappa_*}(e^{\kappa_* v_{*1}}-1)\int^\infty_{v_{*1}}e^{-\kappa_* v_*}
\left(|\sigma_{(1)}|^2+\frac{1}{2}{\cal T}_{(2)} \right)dv_*
\nonumber \\
&\qquad+\frac{2\epsilon^2}{\kappa_*}\int^{v_{*1}}_0 (1-e^{-\kappa_* v_*})
\left(|\sigma_{(1)}|^2+\frac{1}{2}{\cal T}_{(2)} \right)dv_*. 
\end{align}
The first teleological term in the r.~h.~s.  necessarily appears because a black hole is defined 
as a region from which light cannot escape to infinity. So, following the Hawking-Hartle formula in Ref.~\cite{Chandra}, 
we shall assume that  
\begin{align}
|\sigma_{(1)}|^2+\frac{1}{2}{\cal T}_{(2)}=0 \quad \mbox{for} \quad v_*>v_{*1}. 
\end{align}
Thus, by setting $\Sigma=\Sigma_i+\delta \Sigma$, we have 
\begin{align}
\ln\left[\frac{\Sigma_i+\delta \Sigma}{\Sigma_i} \right]\simeq \frac{\delta \Sigma}{\Sigma_i}
=\frac{2\epsilon^2}{\kappa_*}\int^{v_{*1}}_0 \left(|\sigma_{(1)}|^2+\frac{1}{2}
{\cal T}_{(2)} \right)dv_*,  
\end{align}
where we assumed that $v_{*1}\gg \kappa_*^{-1}$. 
This immediately yields Eq.~(\ref{area_variation}).

\end{document}